\documentclass[prb,twocolumn]{revtex4}

\usepackage{graphicx}
\usepackage{dcolumn}
\usepackage{bm}

\begin{document}

\preprint{APS/123-QED}

\title{Quantum Kelvin-Helmholtz instability in phase-separated two-component Bose-Einstein condensates}

\author{Hiromitsu Takeuchi,${}^1$ Naoya Suzuki,${}^2$ Kenichi Kasamatsu,${}^3$ Hiroki Saito,${}^2$  and Makoto Tsubota${}^1$}
\affiliation{%
${}^1$Department of Physics, Osaka City University, Sumiyoshi-ku, Osaka 558-8585, Japan \\
${}^2$Department of Applied Physics and Chemistry, University of Electro-Communications, Tokyo 182-8585, Japan\\
${}^3$Department of Physics, Kinki University, Higashi-Osaka, Osaka 577-8502, Japan
}%


\date{\today}

\begin{abstract}
 We theoretically study the Kelvin-Helmholtz instability in phase-separated two-component Bose-Einstein condensates using the Gross-Pitaevskii and Bogoliubov-de Gennes models.
 A flat interface between the two condensates is shown to deform into sawtooth or Stokes-type waves,
 leading to the formation of singly quantized vortices on the peaks and troughs of the waves.
 This scenario of interface instability in quantum fluids is quite different from that in classical fluids.
\end{abstract}


\maketitle

\section{introduction}

 A vortex sheet exists on the interface layer between two fluids with a relative velocity.
 When the relative velocity is sufficiently large,
 the vortex sheet becomes dynamically unstable and the interface modes are amplified,
 which typically leads to roll-up patterns in the nonlinear stage \cite{Kundu}.
 This shear-flow instability is called the Kelvin-Helmholtz instability (KHI)
 and is related to several familiar phenomena such as wind-generated ocean waves, flapping flags, billow clouds, and sand dunes.
 KHI is a fundamental instability in fluid dynamics
 and appears at various scales ranging from the laboratory to astronomical scales,
 e.g. in plasmatic flows around the earth \cite{plasma} and relativistic flows in astrophysical jets \cite{jet}.

Recently, KHI in quantum fluids, which we refer to as {\it quantum KHI}, has attracted growing interest.
Quantum KHI was first realized by the Helsinki group \cite{Blaauwgeers},
 in which quantized vortices penetrate from the A phase to the B phase of superfluid $^3$He in a rotating cryostat.
 Quantum KHI is also a candidate mechanism of the pulsar glitches of rotating neutron stars with nucleon superflows \cite{neutron},
 the instability of the crystal-superfluid interface in $^4$He \cite{Abe},
 and the vortex formation in atomic Bose-Einstein condensates (BECs) \cite{Henn}.
 A significant difference of quantum fluids from classical fluids is vortex quantization, and hence quantum KHI is expected to yield novel nonlinear dynamics, which have not yet been explored.

The quantum KHI observed in the Helsinki experiment \cite{Blaauwgeers} can be explained by the hydrodynamic theory \cite{Volovik}.
There exist two mechanisms that can cause quantum KHI.
The first is dynamic instability (DI) induced by interface modes (ripplons) with complex frequencies, 
which is a direct analogue of the classical KHI.
 The second is the Landau (thermodynamic) instability (LI) caused by excitation of ripplons with negative energies (frequencies),
 which is absent in classical KHI and thus unique to quantum KHI.
 We call these two mechanisms dynamic and thermodynamic KHI, respectively.
In the helium experiment with dissipation caused by interaction with the normal component and the container wall,
 the thermodynamic KHI precedes the dynamic KHI because the critical velocity for the LI is generally lower than that for the DI \cite{Volovik}.
 In fact, the instability observed in Ref. \cite{Blaauwgeers} was actually thermodynamic KHI,
 in which the number of vortices was detected by NMR.
However, the explicit nonlinear dynamics of vortex nucleation at the interface is unknown because of difficulties in directly observing vortex dynamics and in its hydrodynamic analysis.

\begin{figure*} [hbpt] \centering
  \includegraphics[width=.99 \linewidth]{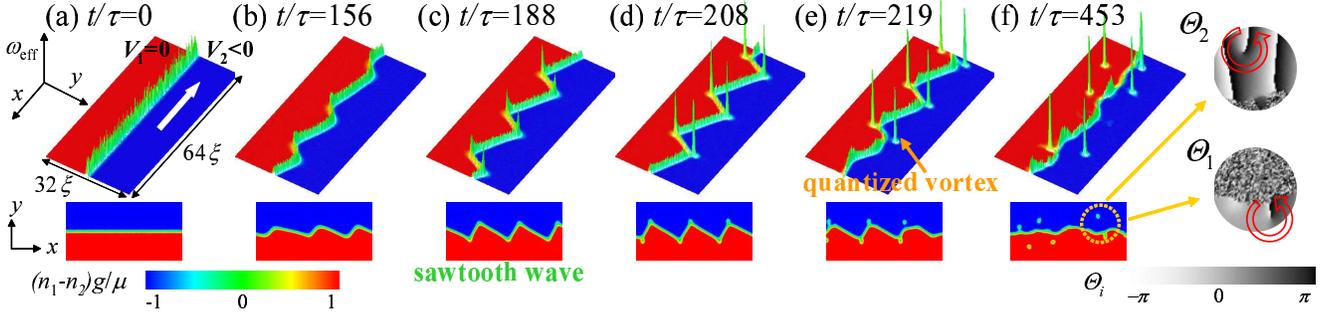}
  \caption{
(color online)
 Nonlinear dynamics of the dynamic KHI in phase-separated two-component BECs without dissipation for the relative velocity $V_{\rm r}=V_1-V_2=0.98c>V_{\rm D}$.
 (Upper panels) The height and color show the vorticity $\omega_{\rm eff}$ and the density difference $n_1-n_2$ between the two condensates, respectively.
 (Lower panels) Two-dimensional plots of $n_1-n_2$. (Right panels) The phases $\Theta_j$ ($j=1,2$) of the condensates for (f).
 The circular arrows show the rotational direction of the superflow around each vortex. 
}
\label{fig:dynamics_N}
\end{figure*}

In this paper, we address the nonlinear dynamics by considering the shear-flow states in phase-separated two-component BECs of ultracold atoms. 
Atomic BECs are ideal systems to examine quantum KHI,
 since we can prepare an interface and shear-flow in a well-controlled manner,
 directly observe the vortex dynamics,
 and realize both dynamic KHI and thermodynamic KHI because of the low dissipation rate.
 We employ the Gross-Pitaevskii (GP) and Bogoliubov-de Gennes (BdG) models, which are more fundamental approaches than the hydrodynamic model in superfluid helium, 
 for studying the stability of the shear-flow states and the vortex dynamics.
 We find that the nonlinear dynamics of quantum KHI are quite different from those for classical KHI.
 Figure~\ref{fig:dynamics_N} shows the typical dynamics of the quantum KHI.
 The initial state exhibits a flat vortex sheet on the interface between the two condensates [Fig. \ref{fig:dynamics_N}~(a)].
 When the relative velocity exceeds a critical value,
 the interface modes are amplified to form sawtooth waves,
 with the vorticity being extremely localized at the peaks and troughs of the waves [Fig.~\ref{fig:dynamics_N}~(c)].
 Then, singly quantized vortices are released from the vortex sheet into the bulk of each condensate [Figs.~\ref{fig:dynamics_N}~(d)--(f)].
 Depending on the relative velocity and dissipation, we find a variety of nonlinear dynamics of quantum KHI.
We also find that quantum KHI is observable in trapped BECs.

\section{shear-flow state in two-component Bose-Einstein condensates}

 We consider two-component BECs in a quasi-two-dimensional system under an external potential $U_j(y)~(j=1,2)$ for component $j$.
 In the mean-field theory at low temperatures,
 the two BECs are described by the condensate wave functions $\Psi_j(t,{\bf r})=\sqrt{n_j(t,{\bf r})}e^{i\Theta_j(t,{\bf r})}$ with the particle densities $n_j$ and the phases $\Theta_j$, which obey the coupled GP equations \cite{Pethick}
\begin{eqnarray}
i \hbar \frac{\partial}{\partial t} \Psi_j = \left(-\frac{\hbar^2}{2m_j}{\bf \nabla}^2+U_j+\sum_{k} g_{jk}|\Psi_k|^2\right)\Psi_j.
\label{eq:GP}
\end{eqnarray}
 Here, $m_j$ is the atomic mass and the coupling constants $g_{11}$, $g_{22}$, and $g_{12}=g_{21}$ are related to the s-wave scattering lengths $a_{11}$, $a_{22}$, and $a_{12}=a_{21}$, respectively, by $g_{ij}=2\pi\hbar^2a_{ij}(m_i+m_j)/m_im_j~(i,j=1,2)$.
 In the stationary state, we suppose that the two components undergo phase separation and form an interface layer along $y=0$,
 sustained by the linear potential $U_j=f_jy$ with a slight inclination $f_j$.
 The condition for the phase separation is given by $\beta=g_{12}/\sqrt{g_{11}g_{22}} >1 $,
 and the densities $n_1$ and $n_2$ are almost zero in the regions $y>0$ and $y<0$, respectively, far from the layer.
 We consider here a sufficiently large value $\beta=10$;
 the interface layer is thick and becomes indistinct when $\beta \sim 1$.
 Each component flows along the $x$-axis with the velocity $V_j$;
 the stationary wave function has a form $\Psi_j^0=\sqrt{\frac{\mu_j}{g_{jj}}}\psi_j(y)e^{i\Theta_j^0}$,
 where $\psi_j$ is a real function and $\Theta_j^0=-\frac{\mu_j}{\hbar}t+\frac{m_j}{\hbar}V_jx$ with the chemical potential $\mu_j$.
 For simplicity, we set the parameters as $\mu=\mu_1-\frac{1}{2}m_1V_1^2=\mu_2-\frac{1}{2}m_2V_2^2$, $g=g_{11}=g_{22}$, $m=m_1=m_2$, and $f=f_1=-f_2$.
 The units of length, time, and velocity are  $\xi=\sqrt{\hbar^2/(m\mu)}$, $\tau=\hbar/\mu$, and $c=\sqrt{\mu/m}$, respectively.

\begin{figure} [htpb] \centering
  \includegraphics[width=.99 \linewidth]{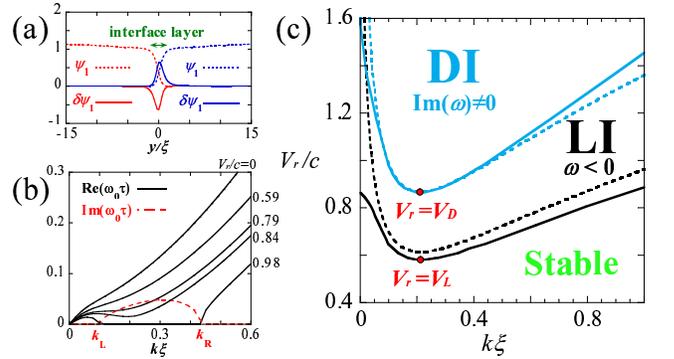}
  \caption{
(color online) (a) Profiles of the wave functions of the two condensates ($\psi_j$) and the excitation ($\delta\psi_j\equiv u_j-v_j^*$) in a shear-flow state.
The real functions $\delta\psi_j$ for the lowest excitation with $k=0.29/\xi$ for $V_{\rm r}=0.79c$ are localized around the interface layer.
 (b) Dispersion relation $\omega_0=\omega-V_{\rm eff}k$ of the lowest modes.
 The broken and solid curves show ${\rm Im}[\omega_0(k)]$ for $V_{\rm r}/c=0.98$ and ${\rm Re}[\omega_0(k)]$ for $V_{\rm r}/c=0$, $0.59$, $0.79$, $0.84$, and $0.98$, respectively. 
  (c) Phase diagram of the quantum KHI for $V_1=0$ and $V_2<0$.
 The  curves are the boundaries of the DI and LI regions obtained by the BdG model (solid curves) and the hydrodynamic model (broken curves).
}
\label{fig:phase_diagram}
\end{figure}

 Figure \ref{fig:phase_diagram}~(a) shows the profile of $\psi_j(y)$ for $f=0.02 \mu/\xi$.
 The shear-flow states are characterized by the effective superflow velocity ${\bf v}_{\rm eff}=(m_1n_1{\bf v}_1+m_2n_2{\bf v}_2)/(m_1n_1+m_2n_2)$ with ${\bf v}_j=\frac{\hbar}{m_j}{\bf \nabla}\Theta_j$.
 The velocity field ${\bf v}_{\rm eff}({\bf r})$ changes sharply at the interface from $V_1\hat{\bf x}$ ($y<0$) to $V_2\hat{\bf x}$ ($y>0$).
 The effective vorticity ${\bf \omega}_{\rm eff}=({\bf \nabla}\times {\bf v}_{\rm eff})_z$ is localized on the interface, which constitutes a vortex sheet.

\section{Linear stability analysis}
\subsection{Hydrodynamic model}

 First, it is instructive to discuss the stability of the shear-flow state with a hydrodynamic model.
 In this model, the interface layer is represented by a curve $y=\eta(t,x)$,
 neglecting its thickness.
 The Lagrangian $L=\int dx \int dy\left( {\cal P}_1+{\cal P}_2-g_{12}n_1n_2 \right)$ of the two-component BECs is approximated with the interface-tension coefficient $\alpha$ and the interface area ${\cal S}$ as
\begin{eqnarray}
L\approx
\int dx
\left(
 \int_{-\infty}^{\eta}dy{\cal P}_1
+\int^{\infty}_{\eta}dy{\cal P}_2
\right)
-\alpha{\cal S}(\eta),
\label{eq:approxLagrangian}
\end{eqnarray}
where ${\cal P}_j=\frac{i\hbar}{2}\Psi_j^*\partial_t \Psi_j-\frac{i\hbar}{2}\Psi_j\partial_t\Psi_j^*-{\cal E}_j$ with ${\cal E}_j=\frac{\hbar^2}{2m_j}|{\bf \nabla}\Psi_j|^2+U_j(y)n_j+\frac{1}{2}g_{jj}n_j^2$.
 The interface tension is obtained by calculating the excess energy due to the presence of the interface \cite{Schaeybroeck}.
 For small $\eta$, we have ${\cal S}(\eta) \approx \int dx \left[1+\frac{1}{2}\left(\partial_x \eta\right)^2\right]$.
 Then the variation in Eq. (\ref{eq:approxLagrangian}) with respect to $\eta$ gives the equation
${\cal P}_1(\eta)-{\cal P}_2(\eta)+\alpha\partial_x^2 \eta=0$,
 which corresponds to the Bernoulli equation in hydrodynamics.
 The motion of the interface can be determined by solving the Bernoulli equation with the kinematic boundary condition \cite{Kundu}.

 To understand the stability qualitatively, the profiles
 $\Theta_j-\Theta^0_j \propto e^{-(-1)^j ky}\cos(kx-\omega t)$ of the phase fluctuations for the surface modes in a single-component BEC \cite{Khawaja} are applied to each surface of the two condensates.
 As a result, for the interface modes with the form $\eta \propto \sin(kx-\omega t)$, we obtain the dispersion relation,
\begin{eqnarray}
\frac{\omega}{k}=V_{\rm eff}+\frac{1}{\sqrt{\rho_1+\rho_2}}\sqrt{\frac{F+\alpha k^2}{k}-\frac{\rho_1\rho_2}{\rho_1+\rho_2}V_{\rm r}^2},
\label{eq:dispersion}
\end{eqnarray}
 where $k>0$, $\rho_j = m_jn'_j$, and $F=n'_1f_1-n'_2f_2$ with the density $n'_j$ near the interface in the Thomas-Fermi approximation $n'_j\approx [\mu-U_j(0)]/g$.
 The frequency $\omega$ depends on the relative velocity $V_{\rm r}=V_1-V_2$ and the effective superflow velocity $V_{\rm eff} \equiv (\rho_1V_1+\rho_2V_2)/(\rho_1+\rho_2)$ at the interface of the stationary state.
 For our parameters
 we have $\rho_1=\rho_2=m\mu/g$, $F=0.04\mu^2/g\xi$, $V_{\rm eff}=(V_1+V_2)/2$, and $\alpha=\sqrt{2}\frac{\mu^2\xi}{g} \left[ 2\sqrt{2}/3 - 0.514\beta^{-1/4} -2 \left( 0.055\beta^{-3/4} + 0.067 \beta^{-5/4} \right) +\cdots \right] \approx 0.886\frac{\mu^2\xi}{g}$ according to Ref. \cite{Schaeybroeck}.

 For  $V_{\rm r}^2 > 2\frac{\rho_1+\rho_2}{\rho_1\rho_2}\sqrt{F\alpha}$,
 the imaginary part ${\rm Im}(\omega)$ becomes nonzero and the shear-flow states are dynamically unstable against excitation of the interface modes with $k_-<k<k_+$ as in classical KHI, where $k_{\pm}=k_0\pm\sqrt{k_0^2-F/\alpha}$ with $k_0=\frac{\rho_1\rho_2}{2\alpha (\rho_1+\rho_2)}V_{\rm r}^2$.
 It is interesting to note that the dispersion relation of $\omega$ with $V_{\rm eff}=0$ has an inflection point as in the roton curve in superfluid helium,
 and Landau's argument of thermodynamic instability also applies to the present system.
 The effective velocity corresponding to the Landau critical velocity is given by
 $\sqrt{2\sqrt{F\alpha}-\frac{\rho_1\rho_2}{\rho_1+\rho_2}V_{\rm r}^2}/\sqrt{\rho_1+\rho_2}$.
 If $V_{\rm eff}$ exceeds this value, there appear interface modes (ripplons) with negative energy quanta $\hbar \omega<0$ and LI occurs in the presence of dissipation.
This instability is unique to quantum KHI since the shear-flow state cannot exist in thermal equilibrium in classical hydrodynamics.

The interface modes with $k \gtrsim 1/\xi$ are not well described with the hydrodynamic model since the interface thickness $\sim \xi$ is neglected there.
In addition, for $k \to 0$, the phase velocity $\omega/k$ in Eq. (3) diverges to infinity.
This unphysical divergence is caused by neglecting density fluctuation of the condensates in this model,
 which is properly treated below.

\subsection{Bogoliubov-de Gennes model}

 We can investigate the stability of the shear-flow states with the BdG model more precisely than the hydrodynamic model. 
 The BdG model describes excitation modes in the bulk in addition to the interface modes,
 considering all degrees of freedom of the fluctuation.
 The fluctuation is written as a collective excitation $\delta \Psi_j=\Psi_j-\Psi^0_j$ around the shear-flow state $\Psi^0_j$.
 Because of the translational symmetry along the $x$-axis,
 the excitation has a form $\delta \Psi_j(t,{\bf r})=e^{i\Theta^0_j(t,x)}\left[u_j(y)e^{-i\omega t+ikx}-v_j^*(y)e^{i\omega t-ikx}\right]$
 with the norm $ \nu \equiv \sum_j\int(|u_j|^2-|v_j|^2)dy \ge 0$.
 The frequency $\omega(k,V_{\rm eff})=\omega_0(k)+V_{\rm eff}k$ is calculated by solving the BdG equations
\begin{eqnarray}
\hbar \omega_0 {\bf u}
&=&
\hat{\cal H}{\bf u},
\label{eq:reducedBdG}\\
\hat{\cal H}
&=&
\left( 
\begin{array}{cccc}
\hat{h}_1^+ & -g_{11}\psi_1^2 &g_{12}\psi_2\psi_1 & -g_{12}\psi_2\psi_1 \\
g_{11}\psi_1^2 & -\hat{h}_1^- &g_{12}\psi_2\psi_1 & -g_{12}\psi_2\psi_1 \\
g_{12}\psi_1\psi_2 & -g_{12}\psi_1\psi_2 & \hat{h}_2^- & -g_{22}\psi_2^2 \\
g_{12}\psi_1\psi_2 & -g_{12}\psi_1\psi_2 & g_{22}\psi_2^2 & -\hat{h}_2^+ \\
\end{array} 
\right),
\nonumber
\end{eqnarray}
 where ${\bf u}=(u_1, v_1, u_2, v_2)^T$, 
$\hat{h}^{\pm}_j=
 \frac{\hbar^2k^2}{2m}-\frac{\hbar^2}{2m}\frac{d^2}{dy^2}+U_j+\sum_k g_{jk}\psi_k^2+g\psi_j^2-\mu \pm\hbar k\frac{V_r}{2}$.

 Figure \ref{fig:phase_diagram}~(b) shows the numerical results for the dispersion relations $\omega_0(k)$ of the excitations with the lowest energies,
 which correspond to the interface modes since $u_j$ and $v_j$ are localized around the interface [Fig. \ref{fig:phase_diagram}~(a)]. 
 In the present discussion, we restrict ourselves to the lowest mode,
 since the other modes have much larger critical velocities.
 A minimum in the curve ${\rm Re}(\omega_0)$ appears for a finite $V_{\rm r}$ and the curve reaches zero at $V_{\rm r}=V_{\rm D}$.
 When $V_{\rm r}>V_{\rm D}$, the imaginary part ${\rm Im}(\omega_0)$ emerges for $k_{\rm L}<k <k_{\rm R}$, where $k_{\rm L}$ ($k_{\rm R}$) corresponds to $k_-$ ($k_+$) in the hydrodynamic model.
 This behavior is similar to that of the interface modes in the hydrodynamic model.

 For small $k$ the lowest mode spreads over the whole system,
 and thus is not regarded as the interface mode.
 This mode causes the density fluctuation of the condensates in the bulk,
 since the profile $\delta \Psi_j$ is continuously transformed to that of $\psi_j$ for $k \to 0$.
 The $k=0$ mode with zero energy always exists in the BdG model \cite{Pethick}.
 Then, for $k \to 0$, the phase velocity $\omega_0/k$ should be asymptotic to the phonon velocity,
 which is finite in the BdG model [see Fig. \ref{fig:phase_diagram}~(b)].
 Since the phase velocity $\left(\omega-kV_{\rm eff} \right)/k \propto 1/\sqrt{k}$ in Eq. (\ref{eq:dispersion}) diverges,
 the BdG model is more accurate than the hydrodynamic model for $k \to 0$.

 Figure~\ref{fig:phase_diagram}~(c) shows the phase diagram for the DI with  ${\rm Im}[\omega(k)]\neq 0$ and the LI with $\omega(k) <0$ for $V_1=0$ and $V_2<0$, and then $V_{\rm eff}=-V_{\rm r}/2$.
 The phase boundaries obtained from the BdG model (solid curves) agree with those from the hydrodynamic model with Eq. (\ref{eq:dispersion}) (broken curves) except for $k \gtrsim 1/\xi$ and $k \to 0$.
 As $V_{\rm r}$ increases,
 the LI region appears at $V_{\rm r}=V_{\rm L}$,
 and the boundary between the LI and stable regions is given by $\omega(k) = 0$.
 The DI region is located above the LI region in Fig.~\ref{fig:phase_diagram}~(c).
 The left and right boundaries between the DI and LI regions correspond to $k_{\rm L}(V_{\rm r})$ and $k_{\rm R}(V_{\rm r})$ in Fig.~\ref{fig:phase_diagram}~(b) and meet at $V_{\rm r}=V_{\rm D}$.
 Note that the DI and LI regions never overlap.
 This is because the modes in the DI region which have zero norm $\nu=0$ and thus zero energy, do not cause energy dissipation and thus do not cause LI,
 where the energy of the fluctuations is formally written as $\hbar\omega\nu$.

\section{Nonlinear dynamics}

When $V_{\rm r}>V_{\rm D}$ or $V_{\rm r}>V_{\rm L}$, the interface modes are amplified, leading to nonlinear dynamics.
 Note that the thermodynamic KHI triggered by LI is not described by the energy-conserving GP Eq. (\ref{eq:GP}).
Here, we introduce a dissipative model,
 which can describe the nonlinear dynamics triggered by both DI and LI \cite{Vortexformation}.
 The dissipative nonlinear dynamics is obtained by solving the modified GP equation
 \begin{eqnarray}
(i-\gamma) \hbar \frac{\partial}{\partial t} \Psi_j = \left(-\frac{\hbar^2}{2m_j}{\bf \nabla}^2+U_j+\sum_{k} g_{jk}|\Psi_k|^2\right)\Psi_j,
\label{eq:GP_dis}
\end{eqnarray}
where $\gamma > 0$ is the phenomenological dissipation constant \cite{Vortexformation}.
 In the linear stage of the instability,
 the DI and LI modes are exponentially amplified as $\exp[|{\rm Im}(\omega)|t]$ and $\exp[\gamma|\omega| t]$, respectively.
 Thus the dynamics crucially depend on $\gamma$ and $V_{\rm r}$.

\subsection{Dynamic KHI}

 We now demonstrate the dynamic KHI with  $\gamma=0$ for $V_{\rm r}=0.98c>V_{\rm D}$,
 where LI is absent because of the absence of dissipation.
  We numerically solve the GP equation in a periodic system with a period $L=64\xi$ along the $x$-axis.
 The initial state is $\Psi^0_j$ with a small random seed to trigger the instability [Fig.~\ref{fig:dynamics_N}~(a)].
 In the linear stage of the instability, the sine wave with $k\xi \sim 0.29$,
 which has the maximum imaginary part $\max_k\left\{{\rm Im}\left[\omega(k)\right]\right\}$, is dominantly amplified.
 As the amplitude becomes large,
 the sine wave is distorted by nonlinearity [Fig.~\ref{fig:dynamics_N}~(b)],
 and deforms into a sawtooth wave [Fig.~\ref{fig:dynamics_N}~(c)].
 We note that the vorticity $\omega_{\rm eff}$ increases on the edges of the sawtooth waves and creates singular peaks [Fig.~\ref{fig:dynamics_N}~(d)].
 Subsequently, each singular peak is released into each bulk,
 becoming a singly quantized vortex with a circulation of $\kappa=h/m$ [Fig.~\ref{fig:dynamics_N}~(e)].

 The release of vortices reduces the vorticity of the vortex sheet and therefore reduces the relative velocity across the interface.
 The relative velocity can be roughly estimated with the total vorticity of the vortex sheet divided by the length of the sheet.
 Since six singly quantized vortices are released from the edges of the sawtooth waves [Fig.~\ref{fig:dynamics_N}~(f)],
 the relative velocity decreases by about $6\kappa/L\sim 0.6 c$ to below the threshold $V_{\rm D}$ for DI,
so that additional vortex creation is suppressed.
 The released vortices drift along the interface and the system never recovers the initial flat interface.
 These nonlinear dynamics are quite different from those in classical KHI,
 where the interface wave grows into roll-up patterns \cite{Kundu}.

\subsection{Thermodynamic KHI}

\begin{figure} [hbtp] \centering
  \includegraphics[width=.99 \linewidth]{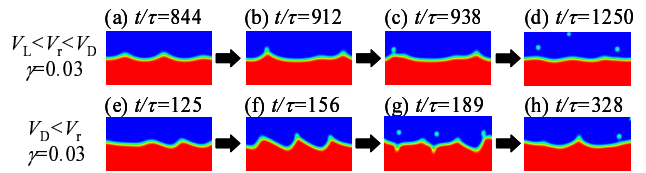}
  \caption{
(color online)
 Dynamics of the density difference $n_1-n_2$ in the quantum KHI with dissipation $\gamma=0.03$ for $V_{\rm L}<V_{\rm r}=0.79c<V_{\rm D}$ (top) and $V_{\rm r}=0.98c>V_{\rm D}$ (bottom).
 The field of view is $32\xi\times64\xi$.
}
\label{fig:dynamics_D}
\end{figure}

 We next consider the thermodynamic KHI with dissipation $\gamma=0.03$ for $V_{\rm L}<V_{\rm r}=0.79c<V_{\rm D}$ [see Fig. \ref{fig:dynamics_D} (top)], where DI is absent.
In the nonlinear stage, we find that the interface has flattened troughs and peaked crests [Fig. \ref{fig:dynamics_D}~(a)] like the Stokes wave known as a finite-amplitude wave at the surface of water \cite{Kundu}.
In this case, the vorticity is highly localized at the crests [Fig. \ref{fig:dynamics_D}~(b)]
 and a singly-quantized vortex is released from each crest into the upper side [Fig. \ref{fig:dynamics_D}~(c)],
 in contrast to the case of the dynamic KHI where the vortices are nucleated in the both sides.

 The dissipation drags the released vortices away from the interface to slow down the upper phase [Fig. \ref{fig:dynamics_D}~(d)].
 Such a slowdown due to the release of quantized vortices is interpreted as phase slippage \cite{Avenel}.
 As a result, four singly quantized vortices are released,
 which reduces the relative velocity across the interface by about $4\kappa/L\sim 0.4 c$ to below the threshold $V_{\rm L}$ for LI.
 Then  the vortex nucleation stops and the vortex sheet recovers a uniform line with less vorticity than that of the initial vortex sheet.

\subsection{Coexistence of dynamic and thermodynamic KHIs}

 For $V_{\rm r}=0.98c > V_{\rm D}$ with dissipation $\gamma=0.03$,
  the dynamic and thermodynamic KHIs coexist.
  Figure \ref{fig:dynamics_D} (bottom) shows typical dynamics after the linear development,
 in which the growth rate $|{\rm Im}(\omega)|$ of the DI is much larger than the rate $\gamma|\omega|$ of the LI.
 In the early stage [Fig.~\ref{fig:dynamics_D}~(e)],
 the dominant DI causes modulation of the interface as in Fig.~\ref{fig:dynamics_N}~(b),
 but the dissipation deforms the sawtooth wave [Fig.~\ref{fig:dynamics_D}~(f)].

 Depending on the wave number of the dominantly amplified mode,
 three vortices are released upward from the peaks of the wave as shown in Fig.~\ref{fig:dynamics_D}~(g).
 As in Figs.~\ref{fig:dynamics_D}~(a)-(d), the vortices released from the interface move upward due to the dissipation.
 In this case, the relative velocity across the interface decreases by $3\kappa/L \sim 0.3 c$ between the thresholds $V_{\rm D}$ and $V_{\rm L}$.
Then the dynamic instability stops and the thermodynamic instability becomes dominant,
 forming the Stokes-type waves [Fig.~\ref{fig:dynamics_D}~(h)].
 The subsequent dynamics are similar to that in Figs. \ref{fig:dynamics_D} (b)--(d).

\subsection{Quantum KHI in trapped condensates}

\begin{figure} [thbp] \centering
  \includegraphics[width=.99 \linewidth]{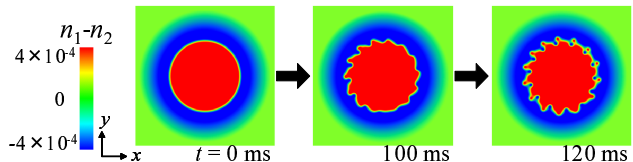}
  \caption{
(color online)
 Dynamics of the density difference $n_1-n_2$ in the quantum KHI without dissipation in a trapped system.
 The total number of atoms is $3.1\times10^6$ with an equal population in each component.
 The trap potential is $U_j({\bf r})=\frac{m}{2}(\omega_{\bot}^2r^2+\omega_z z^2)$ with $r^2=x^2+y^2$, $\omega_{\bot}=2\pi \times 80 {\rm Hz}$, and $\omega_z=2\pi \times 4 {\rm kHz}$,
 and the simulation is performed in quasi-two dimensions.
 The field of view is $72 {\rm \mu m}\times 72{\rm \mu m}$.
}
\label{fig:trap}
\end{figure}
 Experimentally, quantum KHI can be realized in a trapped system.
 We assume that the components $1$ and $2$ are the hyperfine states $|F=1,m_F=-1\rangle$ and $|2,1\rangle$ of $^{87}$Rb atoms, where $a_{11}:a_{12}:a_{22}=0.97:1:1.03$ with their average being $5.5$ nm
 according to Refs. \cite{Matthews, Hall}. 
 Figure~\ref{fig:trap} shows the dynamics of the condensates obtained by numerically solving the GP equation without dissipation,
 where the component $2$ with $30$ vortices surrounds the component $1$ with no vortices in the initial state.
 Such an initial state may be prepared by using a Laguerre-Gaussian beam with orbital angular momentum \cite{Andersen} or the vortex pump proposed in Ref. \cite{Mottonen}.
 The role of the force $F$ in Eq. (\ref{eq:dispersion}) is played by the centrifugal force for the rotating component $2$.
 We see that the dynamic KHI appears on the circular interface.

\section{Conclusions}
 We have studied quantum KHI in phase-separated two-component BECs with shear-flows using the GP and BdG models
and found that a variety of interface patterns are formed,
 such as sawtooth and Stokes-type waves,
 and that quantized vortices make the nonlinear dynamics distinct from those in classical fluids.
 The study of quantum KHI in three-dimensional systems is an interesting direction for future work,
 where the vortex reconnections would change the nonlinear dynamics qualitatively from those in two-dimensional systems.
 Details of quantum KHI in trapped BECs will be published elsewhere.
 We expect that various instabilities in classical fluid mechanics may have counterparts in quantum fluids, in which quantized vortices could yield novel nonlinear dynamics.
 
\begin{acknowledgements}

 This work was supported by KAKENHI from JSPS (Grants No. 199748, 21740267, 20540388, and 21340104) and from MEXT (Grants No. 17071005 and 17071008). 

\end{acknowledgements}

\end{document}